\documentclass[amsmath,amssymb,prc,final,showpacs,twocolumn]{revtex4-1}
 
\usepackage{bm,hyphenat,xspace}
\usepackage{graphicx,epsfig}

\newcommand {\mbf}[1]{{\mathbf{#1}}}

\newcommand {\mcu}{\mathcal{U}}

\newcommand{\He}{{}^3\mathrm{He}}
\newcommand{\Hh}{{}^3\mathrm{H}}

\begin{document}

\title {Neutron-${}^3$H scattering above the
four-nucleon breakup threshold}
 
\author{A.~Deltuva} 
\affiliation{Centro de F\'{\i}sica Nuclear da Universidade de Lisboa, 
P-1649-003 Lisboa, Portugal }

\author{A.~C.~Fonseca} 
\affiliation{Centro de F\'{\i}sica Nuclear da Universidade de Lisboa, 
P-1649-003 Lisboa, Portugal }

\received{June 20, 2012}

\begin{abstract}
The four-body equations of Alt, Grassberger and Sandhas are solved
for the neutron-${}^3$H scattering at energies above the
four-nucleon breakup threshold. The accuracy and practical applicability
of the employed complex energy method is significantly improved 
by the use of integration with the special weights.
This allows to obtain fully converged results with
 realistic nuclear interactions. A satisfactory description of the
existing  neutron-${}^3$H elastic scattering data is obtained.
\end{abstract}

 \maketitle


The four-nucleon reactions is an ideal but also highly challenging
field to test few-nucleon interaction models. The problem of
elastic nucleon-trinucleon scattering 
below the inelastic threshold has already been solved
with high accuracy using several {\it ab initio} methods
with realistic nuclear potentials.  These methods include
the hyperspherical harmonics (HH) expansion 
\cite{viviani:01a,kievsky:08a,viviani:fb19},
the Faddeev-Yakubovsky (FY) equations \cite{yakubovsky:67}
for the wave function components in the coordinate space 
\cite{lazauskas:04a,lazauskas:09a}, and
the Alt, Grassberger and Sandhas (AGS) equations 
\cite{grassberger:67,fonseca:87} for the
transition operators in the momentum space
\cite{deltuva:07a,deltuva:07b,deltuva:08a}.
A recent benchmark \cite{viviani:11a} reported a good agreement between
the HH, FY, and AGS techniques for the
neutron-${}^3$H ($n$-${}^3$H) and proton-${}^3$He ($p$-${}^3$He) scattering. 
Furthermore, deuteron-deuteron ($d$-$d$) collisions, including  the transfer 
reactions to $p$-${}^3$H and $n$-${}^3$He final states,
have been calculated using the resonating-group method (RGM)
\cite{hofmann:07a} and the AGS framework  
\cite{deltuva:07c,deltuva:10a}. However, also these calculations were
 limited  to energies below the three-cluster breakup threshold.
At higher energies, especially above the four-body  breakup threshold,
the asymptotic boundary conditions in the coordinate space
become nontrivial due to open two-, three- and four-cluster channels.
In the momentum-space framework one is faced with a very complicated
structure of singularities in the kernel of integral equations.
Formally, these difficulties can be avoided by rotation
to complex coordinates \cite{lazauskas:11a}
or continuation to complex energy \cite{efros:94a,kamada:03a}
 that lead to bound-state like  boundary conditions 
and nonsingular kernels. However, technical complications may arise
in practical calculations. Indeed,
the applications to the four-nucleon scattering so far have been very
limited \cite{uzu:03a,lazauskas:pc} and none of them 
uses realistic interactions. On the other hand, 
the no-core shell model RGM \cite{quaglioni:08a}
includes in the model space only the ground state of the three-nucleon
system which is insufficient. In Ref.~\cite{frenje:11a}
this shortcoming was partially corrected by adjusting the predictions to 
the experimental data.

The aim of the present work is to overcome the above limitations by
performing realistic well-converged four-nucleon scattering 
calculations above the four-body breakup threshold. We  use
the complex energy method \cite{kamada:03a} but introduce important
technical improvements. Although in the AGS framework employed by
us the Coulomb force can be included via the
screening and renormalization method \cite{alt:80a,deltuva:05a},
the present numerical results are restricted to the Coulomb-free
$n$-${}^3$H case. 

We treat the nucleons as identical particles in the isospin formalism
and therefore use the AGS equations
for the symmetrized four-particle transition operators $\mcu_{\beta \alpha}$
as derived in Ref.~\cite{deltuva:07a}, i.e.,
\begin{subequations}  \label{eq:AGS}   
\begin{align}  
\mcu_{11}  = {}&  -(G_0 \, t \, G_0)^{-1}  P_{34} -
P_{34} U_1 G_0 \, t \, G_0 \, \mcu_{11}  \nonumber \\ 
{}& + U_2   G_0 \, t \, G_0 \, \mcu_{21}, \label{eq:U11}  \\
\label{eq:U21}
\mcu_{21} = {}&  (G_0 \, t \, G_0)^{-1}  (1 - P_{34})
+ (1 - P_{34}) U_1 G_0 \, t \, G_0 \, \mcu_{11}, \\
\mcu_{12}  = {}&  (G_0 \, t \, G_0)^{-1} -
P_{34} U_1 G_0 \, t \, G_0 \, \mcu_{12}  + 
U_2   G_0 \, t \, G_0 \, \mcu_{22}, \label{eq:U12}  \\
\label{eq:U22}
\mcu_{22} = {}& (1 - P_{34}) U_1 G_0 \, t \, G_0 \, \mcu_{12}.
\end{align}
\end{subequations}
Here, $\alpha=1$ corresponds to the $3+1$ partition (12,3)4
whereas  $\alpha=2$ corresponds to the $2+2$ partition (12)(34);
there are no other distinct  two-cluster partitions in the system
of four identical particles.
\begin{gather}\label{eq:G0}
G_0 = (Z - H_0)^{-1}
\end{gather} 
is the free resolvent with the complex energy parameter 
$Z = E+ i\varepsilon$ and the free Hamiltonian $H_0$,
\begin{gather} \label{eq:t}
t = v + v G_0 t
\end{gather} 
 is the pair (12) transition matrix derived from the potential $v$, and 
\begin{gather} \label{eq:AGSsub}
U_\alpha =  P_\alpha G_0^{-1} + P_\alpha t\, G_0 \, U_\alpha,
\end{gather}
are the symmetrized 3+1 or 2+2 subsystem transition operators. 
The basis states are antisymmetric under exchange of two particles in the 
subsystem (12) for the $3+1$ partition 
and in (12) and (34) for the $2+2$ partition.
The full antisymmetry of the four-nucleon system is ensured by the 
permutation operators $P_{ab}$ of particles $a$ and $b$ with
$P_1 =  P_{12}\, P_{23} + P_{13}\, P_{23}$ and $P_2 =  P_{13}\, P_{24}$.

The scattering amplitudes for two-cluster reactions
at available energy $E = \epsilon_\alpha + p_\alpha^2/2\mu_\alpha
=  \epsilon_\beta + p_\beta^2/2\mu_\beta$ are obtained
as the on-shell matrix elements 
$  \langle \mbf{p}_{\beta}| T_{\beta\alpha} |\mbf{p}_{\alpha} \rangle
  = S_{\beta\alpha} 
\langle  \phi_{\beta} | \mcu_{\beta\alpha}| \phi_{\alpha} \rangle $
in the limit $\varepsilon \to +0$. Here 
$|\phi_{\alpha} \rangle $ is the Faddeev component of the
asymptotic two-cluster state in the channel $\alpha$, characterized
by the bound state energy $\epsilon_\alpha < 0$,
the relative momentum $\mbf{p}_\alpha$, and the reduced mass
$\mu_\alpha$. Thus, depending on the isospin, $\epsilon_1$
 is the ground state energy of $\He$ or $\Hh$, and
$\epsilon_2$ is twice the  deuteron  energy $\epsilon_d$.
$S_{\beta\alpha}$ are the
symmetrization factors \cite{deltuva:07a}, e.g., $S_{11} = 3$. 
The amplitudes for breakup reactions 
are given by the integrals involving $\mcu_{\beta\alpha}|\phi_{\alpha} \rangle$
\cite{deltuva:12a}.

We solve the AGS equations \eqref{eq:AGS} in the momentum-space
partial-wave framework. 
The states of the  total angular momentum  $\mathcal{J}$ 
with the projection  $\mathcal{M}$ are defined as  
$ | k_x \, k_y \, k_z   
[l_z (\{l_y [(l_x S_x)j_x \, s_y]S_y \} J_y s_z ) S_z] \,\mathcal{JM} \rangle$ 
for the $3+1$ configuration and 
$|k_x \, k_y \, k_z  (l_z  \{ (l_x S_x)j_x\, [l_y (s_y s_z)S_y] j_y \} S_z)
\mathcal{ J M} \rangle $ for the $2+2$.
Here  $k_x , \, k_y$ and $k_z$ are the four-particle Jacobi momenta
in the convention of Ref.~\cite{deltuva:12a}, 
$l_x$, $l_y$, and $l_z$ are the associated orbital angular momenta,
$j_x$ and $j_y$ are the total angular momenta of pairs (12) and (34),
$J_y$ is the total angular momentum of the (123) subsystem,
 $s_y$ and $s_z$ are the spins of nucleons 3 and 4, 
and $S_x$, $S_y$, and $S_z$ are channel spins
of two-, three-, and four-particle system.
A similar coupling scheme is used for the isospin.
In the following we abbreviate all discrete quantum numbers by $\nu$.
The reduced masses associated with Jacobi momenta $k_x$ and $k_y$
in the partition $\alpha$ will be denoted by $\mu_{\alpha x}$
and $\mu_{\alpha y}$, respectively.

An explicit form of integral equations is obtained by 
inserting the respective completeness relations
\begin{equation} \label{eq:k1k}
1 = \sum_{\nu} \int_0^\infty |k_x k_y k_z \nu \rangle_\alpha
k_x^2 dk_x \, k_y^2 dk_y \, k_z^2 dk_z \,
{}_{\alpha}\langle k_x k_y k_z \nu | 
\end{equation}
between all operators in Eqs.~\eqref{eq:AGS}.
The integrals are discretized using Gaussian quadrature rules \cite{press:89a}
turning  Eqs.~\eqref{eq:AGS} into a system of linear equations 
as described in Ref.~\cite{deltuva:07a}.
However, in the limit $\varepsilon \to +0$ needed for the calculation of 
the observables the kernel of the AGS equations contains
integrable singularities.
At $E + i\varepsilon - \epsilon_{\alpha} - k_z^2/2\mu_\alpha  \to 0 $
the subsystem transition operator in the bound state channel 
has the pole 
\begin{gather} \label{eq:Bpole}
G_0 U_\alpha G_0  \to
\frac{P_\alpha |\phi_\alpha \rangle S_{\alpha\alpha} \langle \phi_\alpha | P_\alpha}
{E + i\varepsilon - \epsilon_{\alpha} - k_z^2/2\mu_\alpha} .
\end{gather}
Furthermore, at 
$E + i\varepsilon - \epsilon_{d} - k_y^2/2\mu_{\alpha y} -k_z^2/2\mu_\alpha \to 0$
the two-nucleon transition matrix in the channel with the deuteron
quantum numbers for the pair (12) has the pole 
\begin{gather} \label{eq:tpole}
 t \to  \frac{v |\phi_d \rangle \langle \phi_d | v }
{E + i\varepsilon - \epsilon_{d} - k_y^2/2\mu_{\alpha y} - k_z^2/2\mu_\alpha},
\end{gather}
with $|\phi_d \rangle$ being the pair (12) deuteron wave function.
Finally, the free resolvent \eqref{eq:G0} obviously becomes singular at 
$E + i\varepsilon - k_x^2/2\mu_{\alpha x} - k_y^2/2\mu_{\alpha y} -
k_z^2/2\mu_\alpha \to 0$. 

At energies below the three-cluster threshold only \eqref{eq:Bpole}
singularities are present that in our previous calculations \cite{deltuva:07a}
were treated reliably by the subtraction technique.
However, above the four-body breakup threshold all three kinds of
 singularities are present. Their interplay with permutation operators
and basis transformations leads to a very complicated singularity
structure of the AGS equations. As proposed in Ref.~\cite{kamada:03a},
this difficulty can be 
avoided by performing calculations for a set of finite $\varepsilon > 0$
values where the kernel contains no singularities
and then extrapolating the results to the $\varepsilon \to +0$ limit.
However, this extrapolation is only precise with not too large
$\varepsilon$ values. On the other hand, for small $\varepsilon$
the kernel of the AGS equations, although formally being nonsingular, 
may exhibit a quasi-singular behavior thereby requiring 
dense grids for the numerical integration. This is no problem
in simple model calculations with rank-one separable potential
and very few channels \cite{uzu:03a} where one can use a large number
of grid points. However, in practical calculations with realistic 
potentials and large number of partial waves necessary for the convergence
one has to keep the number of integration grid points possibly small
and therefore a more sophisticated integration method is needed.

We take over from Refs.~\cite{kamada:03a,uzu:03a} the idea 
of the complex energy method and the $\varepsilon \to +0$ extrapolation 
procedure (analytic continuation via continued fraction)
but we introduce an important technical improvement
when calculating $\mcu_{\beta\alpha}$ at finite $\varepsilon $.
We  use the method of special weights for numerical integrations 
involving any of the above-mentioned quasi-singularities, i.e.,
\begin{gather} \label{eq:wspc}
\int_{a}^{b} \frac{f(x)}{x_0^n + iy_0 - x^n} dx \approx 
\sum_{j=1}^{N} f(x_j) w_j(n,x_0,y_0,a,b).
\end{gather}
The quasi-singular factor $(x_0^n + iy_0 - x^n)^{-1}$ is separated
and absorbed into the special integration weights $w_j(n,x_0,y_0,a,b)$. 
The set of $N$ grid points $\{x_j\}$ where the remaining  smooth function
$f(x)$ has to be evaluated is chosen the same as for the standard
Gaussian quadrature. However,
while the standard weights are real \cite{press:89a}, the special ones
$w_j(n,x_0,y_0,a,b)$ are complex. They are chosen such that 
for a set of $N$ test functions $f_j(x)$ the result \eqref{eq:wspc}
is exact. A convenient and reliable choice of $\{f_j(x)\}$
are the $N$ spline functions  $\{S_j(x)\}$ referring to the grid $\{x_j\}$;
their construction and properties are described in 
Refs.~\cite{press:89a,boor:78a,gloeckle:82a}. The corresponding 
special weights are
\begin{gather} \label{eq:wss}
 w_j(n,x_0,y_0,a,b) = 
\int_{a}^{b} \frac{S_j(x)}{x_0^n + iy_0 - x^n} dx,
\end{gather}
where the integration can be performed either analytically or numerically
using sufficiently dense grid.
This choice of special weights guarantees accurate results for 
quasi-singular integrals \eqref{eq:wspc} with any $f(x)$ that can be
accurately approximated by the spline functions  $\{S_j(x)\}$.

In the integrals over the momentum variables one has
 $n=2$,  $a=0$, and $b\to \infty$. For example, 
when solving the Lippmann-Schwinger equation \eqref{eq:t}
the integration variable in Eq.~\eqref{eq:wspc} is the momentum $k_x$ with
 $x_0^2 = 2\mu_{\alpha x}(E - k_y^2/2\mu_{\alpha y} -k_z^2/2\mu_\alpha)$
and $y_0 = 2\mu_{\alpha x}\varepsilon $.
Alternatively, the quasi-singularity can be isolated in a narrower
interval $0 < a < b < \infty$ and treated by special weights only there.

 Other numerical techniques for solving the four-nucleon AGS equations 
are taken over from Ref.~\cite{deltuva:07a}. They include
Pad\'{e} summation \cite{baker:75a} of Neumann series 
for the transition operators $U_\alpha$ and $\mcu_{\beta \alpha}$
using the algorithm of  Ref.~\cite{chmielewski:03a}
and the treatment of permutation operators (basis transformations)
using the spline interpolation.
The specific form of the permutation operators \cite{deltuva:07a}
leads to a second kind of quasi-singular integrals \eqref{eq:wspc}
with  $n=1$, $a=-1$, $b=1$, and the integration variable
 $x =  \hat{\mbf{k}}'_y \cdot \hat{\mbf{k}}_y$
or  $\hat{\mbf{k}}'_z \cdot \hat{\mbf{k}}_z$ being the cosine of the angle
between the respective initial and final momenta.

We note that the above integration method is not sufficient in the
vanishing $\varepsilon $ limit since for $n=1$ and $y_0=0$ the
result of the integral \eqref{eq:wspc} contains the contribution
$f(x_0) \ln[(x_0+1)/(x_0-1)]$ with logarithmic singularities
at $x_0 = \pm 1$. At finite small  $\varepsilon $  the
result of  \eqref{eq:wspc} may exhibit a quasi-singular behavior.
However, since the logarithmic  quasi-singularity is considerably
weaker than the pole quasi-singularity, at not too small  $\varepsilon $
it is sufficient to use the standard integration.

We start by applying the complex energy method with special integration
weights to the $n$-$\Hh$ scattering below the three-cluster
threshold where our previous results \cite{deltuva:07a} 
obtained at real energies using subtraction technique are available
for comparison. In the test calculations at 3.5 and 6.0 MeV neutron 
energy we find a very good agreement between the two methods,
better than 0.05\% for all relevant phase shifts and observables.
The considered $\varepsilon $
values range between 0.2 and 2.0 MeV, and the number of grid points
is not increased as compared to the real-energy calculations \cite{deltuva:07a}.

\begin{figure}[!]
\begin{center}
\includegraphics[scale=0.69]{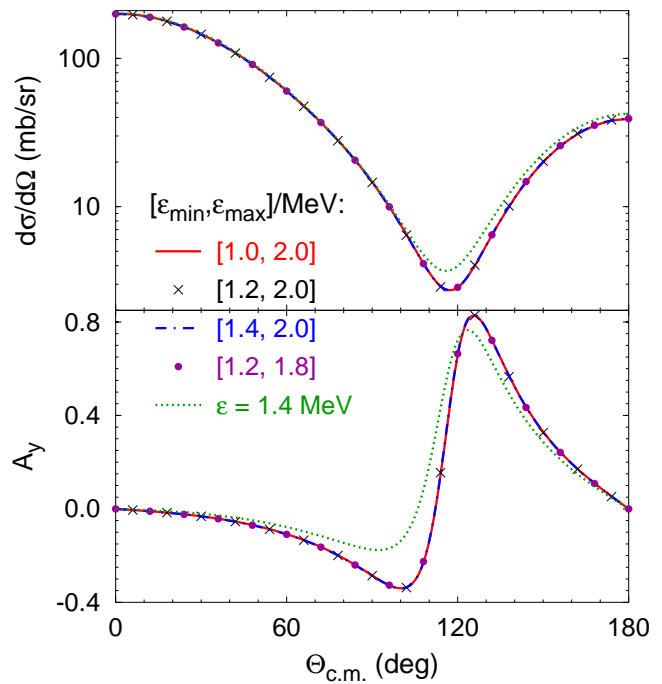}
\end{center} 
\caption{ \label{fig:conv} (Color online)
Differential cross section and neutron analyzing power for
elastic $n$-$\Hh$ scattering at 22.1 MeV neutron energy
as functions of c.m. scattering angle.
Results obtained using different sets of $\varepsilon$ 
values ranging from $\varepsilon_{\mathrm{min}}$ to $\varepsilon_{\mathrm{max}}$ 
with the step of 0.2 MeV are compared; they are indistinguishable.
The dotted curves refer to the  $\varepsilon = 1.4$ MeV calculations
without extrapolation that have no physical meaning but
show the importance of the extrapolation. }
\end{figure}

\begin{table}[!] 
\begin{ruledtabular}
\begin{tabular}{l*{6}{c}}
$[\varepsilon_{\mathrm{min}},\varepsilon_{\mathrm{max}}]$ & 
$\delta ({}^1S_0)$ & $\eta ({}^1S_0)$  & 
$\delta ({}^3P_0)$ & $\eta ({}^3P_0)$  & 
$\delta ({}^3P_2)$ & $\eta ({}^3P_2)$  
\\  \hline
$[1.0,2.0]$ & 62.63 & 0.990 & 43.03 & 0.959 & 65.27 & 0.950 \\
$[1.2,2.0]$ & 62.60 & 0.991 & 43.04 & 0.959 & 65.29 & 0.951 \\
$[1.4,2.0]$ & 62.67 & 0.991 & 43.03 & 0.958 & 65.27 & 0.950 \\
$[1.2,1.8]$ & 62.65 & 0.992 & 43.03 & 0.959 & 65.28 & 0.950 \\
1.4 &         73.37 & 0.916 & 44.77 & 0.840 & 67.38 & 0.933 \\
\end{tabular}
\end{ruledtabular}
\caption{ \label{tab:conv}
Elastic phase shifts (in degrees) and inelasticities in selected 
partial waves for $n$-$\Hh$ scattering at 22.1 MeV neutron energy.
Results for INOY04 potential
obtained using different sets of $\varepsilon$ 
values  ranging from $\varepsilon_{\mathrm{min}}$ 
to $\varepsilon_{\mathrm{max}}$ (in MeV) are compared. In the last line 
the predictions with  $\varepsilon = 1.4$ MeV 
without extrapolation are given.}
\end{table} 

Next we test the numerical reliability of our technique above
the four-nucleon breakup threshold. We use a realistic dynamics,
namely, the high-precision  inside-nonlocal outside-Yukawa
(INOY04) two-nucleon potential  by Doleschall
\cite{doleschall:04a,lazauskas:04a} that 
reproduces experimental binding energies of $\Hh$ (8.48 MeV)
and $\He$ (7.72 MeV) without an irreducible three-nucleon force.
We consider a large number of four-nucleon partial waves sufficient 
for the convergence,
i.e., $l_x,l_y,l_z,j_x,j_y,J_y \le 4$ and $\mathcal{J} \le 5$.
Including more partial waves yields only entirely insignificant changes.
There are too many numerical parameters (numbers of points for various
integration grids) to demonstrate the stability of our calculations
with respect to each of them separately. We found that 10 grid points
are sufficient for all angular integrations but 30 to 40 grid points
are needed for the discretization of Jacobi momenta.
The  $\varepsilon \to +0$ extrapolation yields stable results
only if sufficiently small $\varepsilon$ are considered and 
at each of them the respective calculations are numerically well converged.
We therefore study in Fig.~\ref{fig:conv} the stability of the
 $\varepsilon \to +0$ results obtained via extrapolation
using different  $\varepsilon$ sets ranging from
$\varepsilon_{\mathrm{min}}$ to $\varepsilon_{\mathrm{max}}$ 
with the step of 0.2 MeV. We show the differential cross section
$d\sigma/d\Omega$ and neutron analyzing power $A_y$ for
elastic $n$-$\Hh$ scattering at $E_n = 22.1$ MeV neutron energy.
We find a very good agreement between the results obtained
with $[\varepsilon_{\mathrm{min}},\varepsilon_{\mathrm{max}}] =$ [1.0,2.0],
[1.2,2.0], [1.4,2.0], and [1.2,1.8] MeV, confirming the reliability
of our calculations. 
In addition, we show in Fig.~\ref{fig:conv} the predictions
referring to $\varepsilon = 1.4$ MeV  without extrapolation
that don't have physical meaning. The difference between
 $\varepsilon \to +0$ and $\varepsilon = 1.4$ MeV results
demonstrates the importance of the extrapolation.
Furthermore, in Table \ref{tab:conv} we collect the corresponding 
values for selected phase shifts $\delta$ and inelasticities $\eta$,
i.e., we parametrize the elastic $S$-matrix as $s = \eta e^{2i\delta}$.
As already can be expected from  Fig.~\ref{fig:conv},
the stability of the results with respect to changes in
$[\varepsilon_{\mathrm{min}},\varepsilon_{\mathrm{max}}]$ is very good.
The variations are slightly larger in  the ${}^1S_0$ state 
where also the difference between
the finite  $\varepsilon$ and $\varepsilon \to +0$ results
is most sizable.
Nevertheless, from Fig.~\ref{fig:conv} and  Table \ref{tab:conv}
one can conclude that with a proper $\varepsilon$ choice
 as few as four different $\varepsilon$ values are sufficient to obtain
the physical $\varepsilon \to +0$ results with good accuracy.

For curiosity, in $\mathcal{J}=0$ states 
we performed the calculations keeping the same grids but
with standard integration weights. 
We found that they fail completely at $\varepsilon$ values
from Table \ref{tab:conv}, with the errors of the 
$\varepsilon \to +0$ extrapolation being up to 10 \% for
phase shifts and up to 25 \% for inelasticity parameters.
On the other hand,
at large $\varepsilon > 4$ MeV the two integration methods
agree  well but the $\varepsilon \to +0$  extrapolation 
has at least one order of magnitude larger inaccuracies than those
in Table \ref{tab:conv}.

\begin{figure}[!]
\begin{center}
\includegraphics[scale=0.66]{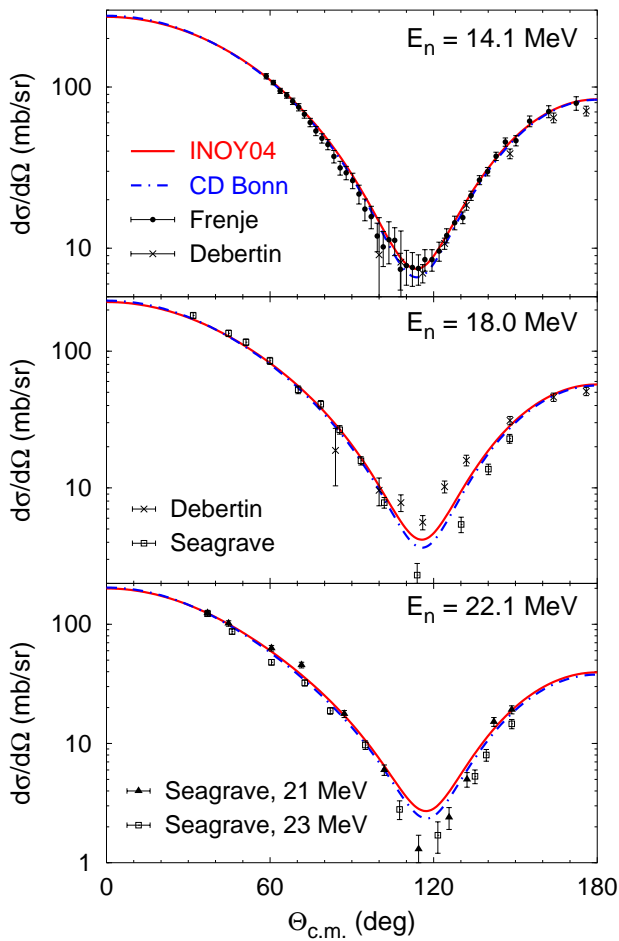}
\end{center} 
\caption{ \label{fig:dcs} (Color online) 
Differential cross section for elastic $n$-$\Hh$ scattering at 
14.1, 18.0, and 22.1 MeV neutron energy. 
Results obtained with INOY04 (solid curves) and CD Bonn (dashed-dotted curves)
potentials are compared with the experimental data from
Refs.~\cite{frenje:11a,debertin:exfor,seagrave:72}.}
\end{figure}

\begin{figure}[!]
\begin{center}
\includegraphics[scale=0.66]{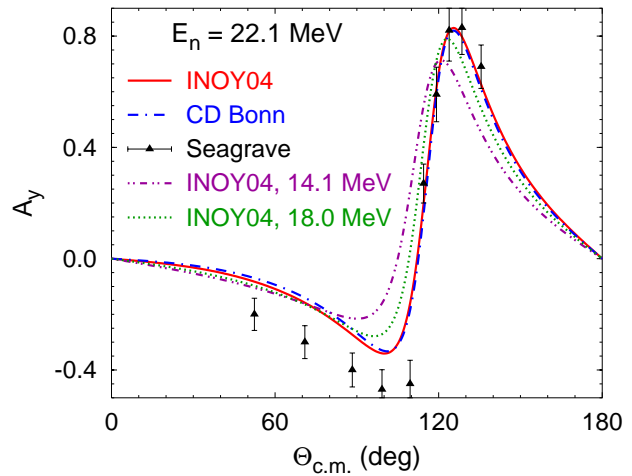}
\end{center} 
\caption{ \label{fig:ay} (Color online) 
Neutron analyzing power for elastic $n$-$\Hh$ scattering.
INOY04  and CD Bonn predictions at $E_n = 22.1$ MeV  are
compared with the data from Ref.~\cite{seagrave:72}. 
INOY04 results at 14.1 MeV and 18.0 MeV are also shown.}
\end{figure}

After establishing the reliability of our calculations we proceed
to the comparison  with the experimental data.
In addition to the INOY04 potential we present results derived
from the CD Bonn potential \cite{machleidt:01a} that
underbinds the $\Hh$ nucleus by 0.48 MeV.
In Fig.~\ref{fig:dcs} we show the differential cross section
for elastic neutron-$\Hh$   scattering at 14.1, 18.0, and 22.1 MeV
neutron energy. Except for the minimum around 115 degrees, the predictions
are insensitive to the choice of the potential.
At $E_n = 14.1$ MeV the new data set by Frenje {\it et al.} \cite{frenje:11a}
is described very well. Other existing data at this energy are
consistent with Ref.~\cite{frenje:11a} but have larger error bars;
we only show the data by Debertin {\it et al.} \cite{debertin:exfor}.
At 18.0 MeV the data sets by Debertin {\it et al.} \cite{debertin:exfor}
and Seagrave  {\it et al.} \cite{seagrave:72} are inconsistent
with each other around the minimum while the theoretical predictions lie
in the middle. The results at $E_n = 22.1$ MeV are compared with 
the data  taken at 21 and 23 MeV by  Seagrave  {\it et al.} \cite{seagrave:72}.
The predictions lie between the two data sets except for the minimum
region. However, given the agreement between  the \cite{frenje:11a}
and \cite{debertin:exfor} data and disagreement between the
\cite{seagrave:72} and \cite{debertin:exfor} data, one may question
the reliability of the data by  Seagrave  {\it et al.} in the minimum region.
Thus, new measurements are needed to resolve this discrepancy.

In  Fig.~\ref{fig:ay} we present the neutron analyzing power for
elastic $n$-$\Hh$ scattering at $E_n = 22.1$ MeV. To study the energy
dependence we also show INOY04 predictions at $E_n = 14.1$ and 18.0 MeV.
We observe that the sensitivity to the nuclear force model  and
energy  is considerably weaker as compared to the regime
below three-cluster threshold \cite{deltuva:07a,deltuva:07b}.
Most remarkably,
in contrast to low energies where the famous $p$-$\He$ $A_y$-puzzle
exists \cite{viviani:01a,fisher:06,deltuva:07b},  the peak of $A_y$
around 120 degrees is described very well but
there is a discrepancy in the minimum region.
This is somehow similar to the three-nucleon system where the
nucleon-deuteron $A_y$-puzzle existing at low energies disappears
as the energy increases \cite{gloeckle:96a}.

In this work we do not calculate explicitly the breakup reactions.
However, the total  $n$-$\Hh$ cross section $\sigma_t = \sigma_e + \sigma_b$
with the elastic $\sigma_e$ and three- and four-cluster breakup $\sigma_b$
contributions is calculated
using the optical theorem. The results at three considered energies
are collected in Table \ref{tab:tcs}. The $\sigma_t$ predictions are slightly
lower than the experimental data from Refs.~\cite{battat:59,phillips:80}
but in most cases they agree within error bars. 
The breakup cross section $\sigma_b$ is most sensitive to 
the potential model, probably due to differences in
$\Hh$ binding energy and breakup threshold positions.

\begin{table}[!]
\begin{ruledtabular}
\begin{tabular}{l*{8}{c}}
&  \multicolumn{3}{c}{INOY04} &
 \multicolumn{3}{c}{CD Bonn} & 
 \multicolumn{2}{c}{Experiment}  \\
$E_n$ & $\sigma_e$ &  $\sigma_b$  & $\sigma_t$  
& $\sigma_e$ & $\sigma_b$ & $\sigma_t$  & $\sigma_t$ 
& Ref. \\
\hline
14.1 & 928 & 19 & 947 & 913 & 28 & 941 & $978\pm 70$ & \cite{battat:59} \\
18.0 & 697 & 41 & 738 & 689 & 48 & 737 & $750\pm 40$ & \cite{battat:59} \\
22.1 & 536 & 61 & 597 & 524 & 70 & 594 & $620\pm 24$ & \cite{phillips:80} \\
\end{tabular}
\end{ruledtabular}
\caption{ $n$-$\Hh$ elastic $\sigma_e$, breakup $\sigma_b$, and total 
$\sigma_t$ cross sections (in mb) at selected neutron energies (in MeV).}
\label{tab:tcs}
\end{table} 

In summary, we performed fully converged neutron-$\Hh$
scattering calculations with realistic potentials
above the four-nucleon breakup threshold. The symmetrized Alt, Grassberger,
and Sandhas four-particle equations were solved in the momentum-space
framework.
We greatly improved the accuracy and the efficiency of the complex
energy method by using numerical integration technique with special
weights for all quasi-singularities of the pole type.
The calculated elastic scattering observables show little sensitivity
to the interaction model. The overall agreement with the 
experimental data for the elastic differential cross section and neutron
analyzing power as well as for the total cross section
is quite good, considerably better than in the low-energy 
$n$-$\Hh$ and  $p$-$\He$ scattering. 
Extension of the method to other reactions in the four-nucleon
system is in progress.

The author thanks R.~Lazauskas and A.~C. Fonseca
for valuable discussions and J.~A. Frenje for providing the data.


\end{document}